\documentclass{bmcart}

\usepackage[utf8]{inputenc} 
\usepackage[section]{placeins}
\usepackage{graphicx}
\usepackage{placeins}
\usepackage{url}

\startlocaldefs
\endlocaldefs

\begin{document}

\begin{frontmatter}

\begin{fmbox}
\dochead{Research}


\title{Impact of Network Centrality and Income on Slowing Infection Spread after Outbreaks}


\author[
   addressref={aff1},                   
   email={yucelshiv@gmail.com}   
]{\inits{SGY}\fnm{Shiv G.} \snm{Yücel}}
\author[
   addressref={aff2},
   email={rafael.pereira@ipea.gov.br}
]{\inits{RHMP}\fnm{Rafael H. M.} \snm{Pereira}}
\author[
   addressref={aff3},
   email={ppeixoto@usp.br}
]{\inits{PSP}\fnm{Pedro S.} \snm{Peixoto}}
\author[
   corref={},                       
   noteref={},                        
   addressref={aff4},
   email={f.camargo@exeter.ac.uk}
]{\inits{CQC}\fnm{Chico Q.} \snm{Camargo}}


\address[id=aff1]{
  \orgname{Oxford Internet Institute, University of Oxford}, 
  \city{Oxford},                              
  \cny{UK}                                    
}
\address[id=aff2]{%
  \orgname{Institute for Applied Economic Research (IPEA)},
  \city{Brasília},
  \cny{Brazil}
}
\address[id=aff3]{%
  \orgname{Applied Mathematics Department, University of São Paulo},
  \city{São Paulo},
  \cny{Brazil}
}
\address[id=aff4]{%
  \orgname{Department of Computer Science, University of Exeter},
  \city{Exeter},
  \cny{UK}
}


\begin{artnotes}
\end{artnotes}

\end{fmbox}


\begin{abstractbox}

\begin{abstract} 
The COVID-19 pandemic has shed light on how the spread of infectious diseases worldwide are importantly shaped  by both human mobility networks and socio-economic factors. Few studies, however, have examined the \emph{interaction} of mobility networks with socio-spatial inequalities to understand the spread of infection.  We introduce a novel methodology, called the Infection Delay Model, to calculate how the arrival time of an infection varies geographically, considering both effective distance-based metrics and differences in regions' capacity to isolate -- a feature associated with socioeconomic inequalities. To illustrate an application of the Infection Delay Model, this paper integrates household travel survey data with cell phone mobility data from the São Paulo metropolitan region to assess the effectiveness of lockdowns to slow the spread of COVID-19. Rather than operating under the assumption that the next pandemic will begin in the same region as the last, the model estimates infection delays under every possible outbreak scenario, allowing for generalizable insights into the effectiveness of interventions to delay a region's first case. The model sheds light on how the effectiveness of lockdowns to slow the spread of disease is influenced by the interaction of  mobility networks and socio-economic levels. We find that a negative relationship emerges between network centrality and the infection delay after lockdown, irrespective of income.  Furthermore, for regions across all income and centrality levels, outbreaks starting in less central locations were more effectively slowed by a lockdown. Using the Infection Delay Model, this paper identifies and quantifies a new dimension of disease risk faced by those most central in a mobility network. 
\end{abstract}


\begin{keyword}
\kwd{human mobility}
\kwd{socio-economic inequality}
\kwd{epidemic intervention effectiveness}
\kwd{spatial analysis}
\end{keyword}


\end{abstractbox}
%

\end{frontmatter}




\section*{Introduction}
Since the start of the COVID-19 pandemic, an active literature has evolved to study the spread and dynamics of the disease from mobility networks \cite{Coelho2020, Peixoto2020, Chang2020, levin} or socio-spatial perspectives \cite{Lee2021, LiSabrinaPereira2020, Cordes2020}. However, very few studies look at how \emph{both} socio-economic conditions and network properties interact, and how they influence outbreaks together. Further, while extensive work has been done to model the spread of the virus and non-pharmaceutical intervention effectiveness in terms of cases, hospitalizations, and deaths \cite{LiSabrinaPereira2020, Cordes2020, Flaxman2020, Oraby2021, Oka, Meo2020} there is a lack of emphasis on the timing of case spread, and how interventions can delay a region's first infection. Given the spatio-temporal granularity of cell phone mobility data capturing responses to lockdown policies, it is now possible to develop generalized, preventative methodologies which seek to further our understanding of disease vulnerability, and better prepare for novel outbreaks or variants.  \par 

This paper develops the Infection Delay Model (IDM), a novel effective distance-based methodology that can be used for assessing how lockdowns can delay a region's first case and their intersection with socio-economic inequalities. The IDM captures the difference between disease arrival times with and without a lockdown, using a novel application of cellphone mobility data for effective distance research. To develop a forward-looking understanding of the impacts of interventions on the timing of disease spread, a use-case of the IDM is presented which considers the potential variability of future outbreak scenarios. Drawing from recent studies of network-driven contagion phenomena \cite{Brockmann2014,Iannelli2017,Balcan2009}, we simulate epidemics from every node in the transport network. By connecting those simulations with socio-economic data, generalizable insights are uncovered which can be applicable beyond the specific spreading patterns observed during COVID-19. 

This paper uses the Metropolitan Region of São Paulo (MRSP) as a case study to apply the IDM. Given its unique position as an area of early disease introduction and high intrastate transmission, COVID-19 studies in the MRSP can help with preparation for future variants of COVID-19 or other pandemics \cite{Coelho2020, Candido2020}. 

\section*{Background}
\subsection*{Network-Based Analyses of COVID-19}
One branch of literature on COVID-19 has focused on mobility networks to model the spread of the disease and assess the risks of cases and deaths. The data sources used to generate such networks range from domestic and international flight records \cite{Coelho2020, Kuo2021}, to cell phone mobility records and geo-located visits to places of interest \cite{Peixoto2020, Chang2020, Sheen2020, Ferreira2021}. The varying spatio-temporal granularity of the data sources used in these analyses have led to diverse outputs to identify regions at risk and explore how non-pharmaceutical interventions (NPIs) such as lockdowns impact mobility and vulnerability.

This area of literature uses transport flows to construct aggregated networks of population movement. Various methods have been implemented to study COVID-19 risks on these mobility networks. Effective distance-based studies calculate the `distance' between two regions based on the degree of mobility flows between them -- more connected regions are effectively `closer' \cite{Coelho2020}. The effective distance of a region from an outbreak location has been shown to be predictive of infection arrival times \cite{Brockmann2014, Iannelli2017}. Other studies build compartmental models on top of the mobility networks, calibrated to regional epidemic trajectories, and use epidemiological parameters and outbreak locations to simulate the course of an epidemic \cite{Chang2020, Balcan2009, Peixoto2020, Sheen2020}. As greater mobility and person-to-person contact is associated with transmission, epidemic simulations can be run on mobility networks with adjusted levels of mobility or contact patterns to explore the impacts of real or hypothetical interventions on health outcomes \cite{levin, Sheen2020}.

\subsection*{Socio-Spatial Analyses of COVID-19}
A separate branch of literature on COVID-19 has focussed on disease vulnerability and its intersection with existing socio-spatial inequalities. The range of analyses includes studies on how socioeconomic levels are associated with differences in terms of cases, hospitalizations, and deaths \cite{LiSabrinaPereira2020, Cordes2020}, health care facility access \cite{Pereira2021, Tao2020} and inequalities in NPI adherence \cite{LiSabrinaPereira2020, Lee2021, Jay2020, Heroy2022}. These spatial analyses often seek to uncover how variables such as race and income relate to COVID-19 risks, to identify how existing inequalities are being compounded by the ongoing pandemic. \par 

In an analysis of hospitalization and deaths in São Paulo, it was found that black and \emph{pardo} Brazilians were more likely to be hospitalized and die of COVID-19 \cite{LiSabrinaPereira2020}. Similarly, an analysis of clusters and contextual factors of COVID-19 in New York City found that regions with larger black populations without health insurance had higher positive testing rates \cite{Cordes2020}. Cell phone mobility data has also been used to study the interaction of lockdown adherence and socio-economic inequalities. Conceptualizing mobility restrictions as a luxury not everyone can afford, it has been found that more vulnerable individuals were less able to reduce their mobility -- potentially due to a lower probability of furlough or teleworking opportunities \cite{Lee2021, LiSabrinaPereira2020}. \par 

\subsection*{Contributions}
The first contribution of this paper is the integrated analysis of mobility networks and socio-economic characteristics to measure disease risk. While the socio-spatial branch of literature has consistently identified intersections between socio-economic vulnerability and disease burdens, current network-based studies either omit socio-economic data, or include it to identify at-risk regions based on network-based results \cite{Coelho2020}. There is a lack of investigation into the interaction between network properties and socio-economic factors, and how they jointly impact the distribution of disease risk. It cannot be assumed that features such as network centrality and income are proxies for each other, justifying an investigation which explicitly examines both. \par 

The second contribution of this study is its focus on the ability of lockdown restrictions to slow the spread of disease. Existing network-based and socio-spatial research on cases, hospitalizations, and deaths fail to measure a crucial goal of early lockdowns -- slowing the spread of disease and delaying the time until a region's first case. Delaying disease onset with early interventions can buy time for health systems to increase hospital and intensive care capacity, and establish rapid testing sites \cite{Rocha2021}. This unexplored dimension of disease risk is investigated using the Infection Delay Model, an effective distance-based method of calculating disease arrival times under baseline and lockdown mobility scenarios. Current literature which explores rankings of disease arrivals using effective distances does so while assuming a single known outbreak location \cite{Brockmann2014, Iannelli2017}, or including a small subset of potential outbreak locations \cite{Coelho2020}. These studies also overlook how differences in social isolation levels across regions are shaped by socioeconomic inequalities. Given recent literature on the outsized influence of the outbreak region on the trajectory of a communicable disease \cite{brockmannoutbreak}, this study simulates outbreaks beginning in every region of the MRSP, to allow for generalizable findings that do not assume that the next outbreak will begin in the same region as the last.

\section*{Data and Interpolation}

\subsection*{Cellphone Mobility Data}
Through an agreement with InLoco \cite{incognia}, a Brazilian cellphone analytics company now known as Incognia, this paper had access to daily isolation levels for MRSP from March 1, 2020 to April 19, 2020. These data come spatially aggregated on a hexagonal grid using the H3 index at resolution 8 \cite{brodsky_2019}. The data set contains 2893 hexagonal cells of roughly 740m$^2$ across the MRSP, of which 2599 had suitable time frames and auxiliary data after interpolation to be used in the analysis. The hexagonal isolation data is openly available in a data repository (see Availability of Data and Materials section). InLoco/Incognia gathers data by partnering with mobile phone applications, and uses software development kits to harvest location data while individuals are using the partnered app \cite{Peixoto2020}. This form of location gathering provides precise geo-coordinates, which are anonymized and aggregated to develop the social-isolation indices. For a given hexagon cell, the proportion of individuals who reside in the cell and \emph{stay} within it on a given day is recorded. This proportional value is used as a proxy for social isolation \cite{LiSabrinaPereira2020}, recording the extent to which individuals travel outside their residence area. Higher or lower social isolation values indicate that \emph{fewer} or \emph{more} individuals are leaving their residence area, respectively \cite{Ferreira2021}. The distribution of social isolation hexagon cells is presented in Figure \ref{figure:pop_ex}. The uneven coverage of the MRSP hexagon cells is a feature of the data set provided by InLoco/Incognia, discussed in the limitations section.

\subsection*{Travel Survey Data}
The travel survey data for the MRSP were gathered from the 2017 MRSP household travel survey, conducted by the São Paulo Metropolitan Transportation Department between June 2017 and October 2018 \cite{sptravel}. The original data set is a table of survey responses regarding the total daily trips of 86,318 individuals who reside in the MRSP. On average, each individual reports 2.12 daily journeys, leading to a total of 182,994 trip reports \cite{sptravel}. Key information for the reports are the journey origin and destination, along with the travel time. The interviews were conducted across 39 municipalities within the MRSP, divided into 510 research zones for the purposes of the survey. Of all the research zones, 66\% lie within the main municipality in the MRSP, São Paulo. The survey was designed to be statistically representative across the MRSP, and includes journey and population weights to scale responses by their frequency in the true population. From these weights, the total 2017 mobility flows between travel survey zones and 2017 estimates of populations were calculated. Population levels in 2020 were estimated by determining the geometric growth rate from 2010 and 2019 population totals, and scaling the 2017 populations \cite{2010pop, 2019pop}.  \par

\subsection*{Census Data}
This study uses socio-economic data from the official 2010 Brazilian Census, focused on the census tracts within the MRSP \cite{census, pereira2019geobr}. Within the state of São Paulo, there are 68,296 tracts, with data included on the total population, racial aggregates, average income per capita (Brazilian Real per calendar month), functioning water networks, and other relevant socio-economic features. The census tracts within São Paulo state cover a larger area than both the cellphone mobility hexagon cells and travel survey zones, which are primarily focused on the MRSP. The population data from the the MRSP travel survey is more up to date than the 2010 census, therefore it is used in favour of the census data population totals. The census data remains useful for calculating regional income per capita averages, which are interpolated from census tracts into the hexagon cells.

\subsection*{Interpolating Data to Hexagon-Level}
 While the social isolation hexagon cells provide spatially and temporally granular information on the daily proportion of residents leaving a given area, information on which population subgroups are included in each hexagon cell remain unknown. This problem is shared across the growing body of literature using cellphone mobility data for public health purposes, where anonymity measures by cellphone data providers obscure information on the sample \cite{Grantz2020}. While fundamental selection biases in the mobile phone data are a persistent issue, discussed in the limitations section, traditional data sources can be leveraged to generate population estimates within the hexagons \cite{Aleta2020}. \par 

\begin{figure}[!h]
    \includegraphics[width=12cm]{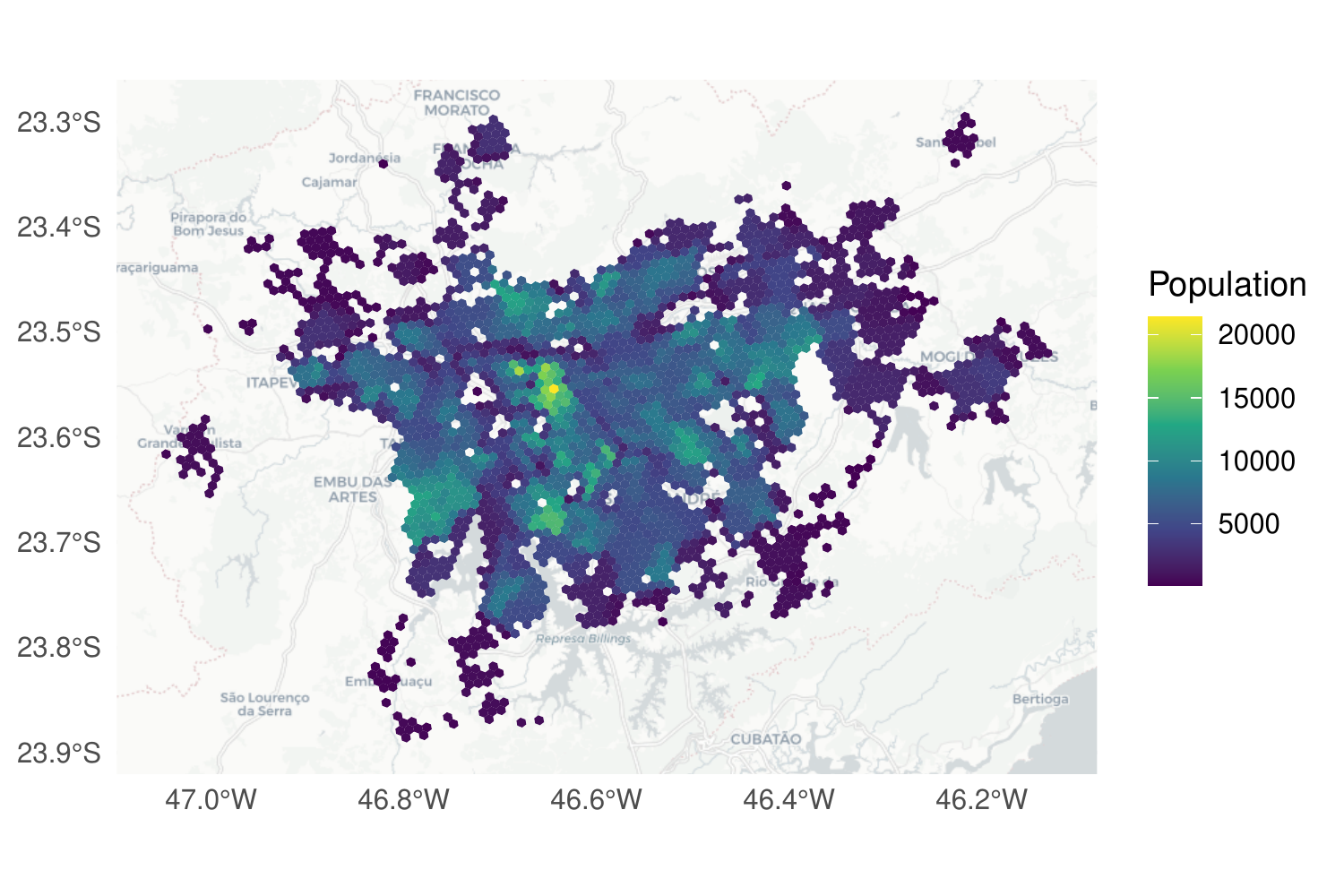}
       \caption{Geographic distribution of all hexagonal cells in the Metropolitan Region of São Paulo for which daily isolation levels are available from the mobile analytics company InLoco/Incognia \cite{incognia}. Shows variation in populations across the Metropolitan Region of São Paulo.}
		\label{figure:pop_ex}
    \end{figure}

\begin{figure}[h!]
    \includegraphics[width=8cm]{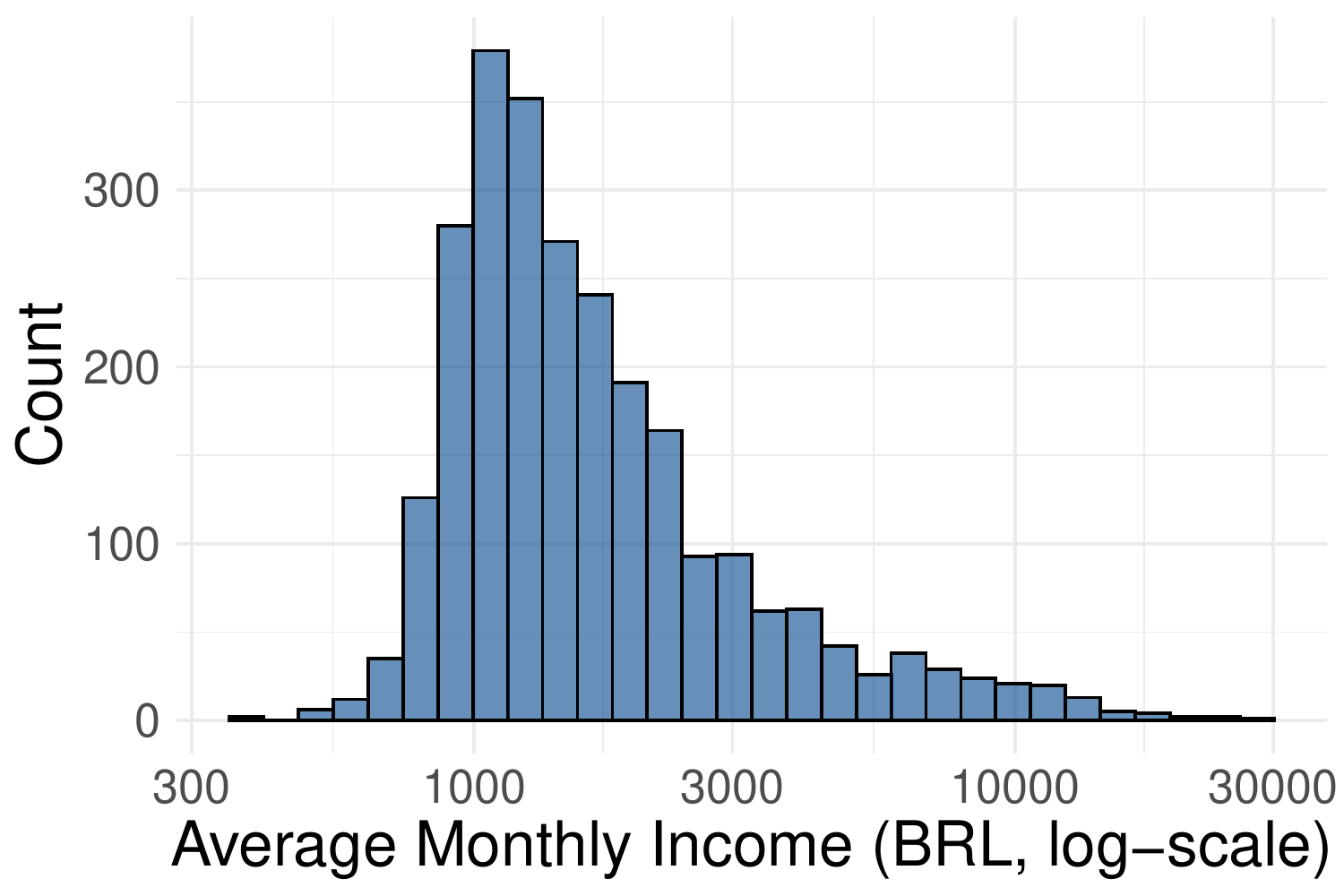}
    \caption{Distribution of average income per capita (Brazilian Real per month) across hexagon cells.}
    \label{figure:income_dist}
\end{figure}

The census tracts and travel survey zones are constructed of varying spatial structures which must be mapped to the social isolation hexagon cells. This process, known as spatial interpolation, is used in geospatial studies to estimate values in unknown area units using values in known geographic units \cite{comber}. The spatial interpolation method used in this analysis is known as aerial weighting, which integrates socio-economic estimates based on proportional overlap \cite{comber}. This method depends on the assumption of homogeneously distributed characteristics within census tracts and travel survey zones, but benefits from transparency and simplicity relative to interpolation methods which rely on auxiliary information \cite{comber}. Each hexagon cell's overlap with census tracts and travel survey zones was determined relative to their total areas. This proportional overlap area was used to generate a weighted allocation for income and population levels. For example, if a hexagon cell covered 50\% of a travel survey zone with a population of 20, the hexagon cell would be assigned 10 individuals. Figure \ref{figure:pop_ex} geographically displays the interpolated populations across the hexagon cells in the MRSP, and Figure \ref{figure:income_dist} and Table \ref{table:income_table} display the distribution of incomes. \par

\FloatBarrier

\FloatBarrier

\begin{table}[!htb]
\begin{tabular}{lr}
\hline
{} &        Monthly Income \\
{} &        (BRL/capita) \\
\hline
mean  &   2172.079 \\
std   &   2318.79\\
min   &      0.00 \\
25\%   &   1069.62 \\
50\%   &   1418.27 \\
75\%   &   2187.37 \\
max   &  28471.98 \\
      \end{tabular}
      \label{table:income_table}
\caption{Summary statistics of average income per capita (Brazilian Real per month) across hexagon cells.}
\end{table}

\begin{figure}[h!]
     \includegraphics[width=12cm]{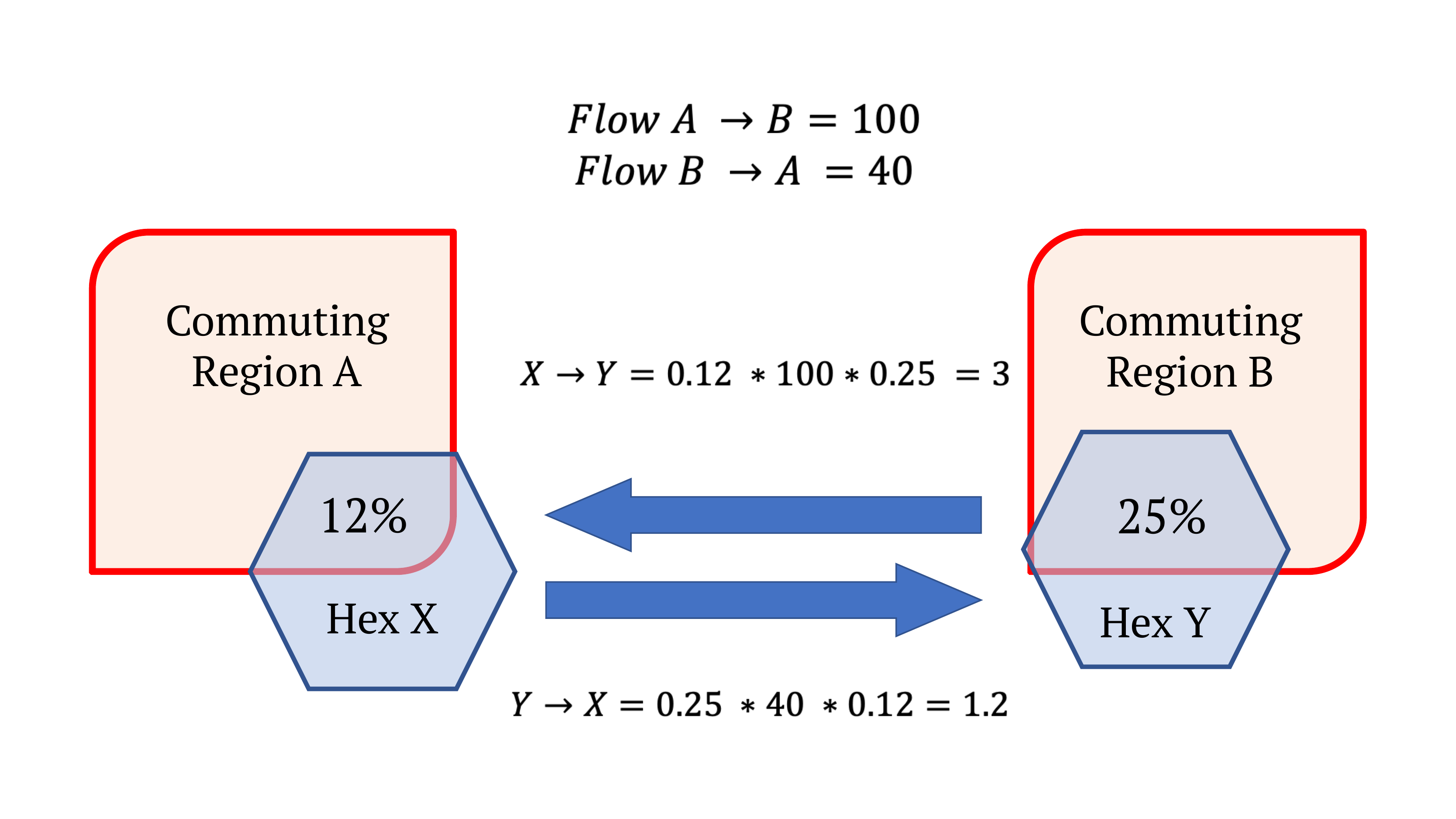}

  \caption{
Travel survey data interpolation strategy from travel zones to hexagon cells. Outflow and inflow are proportional to a hexagon cell's overlap with a travel zone. To estimate travel patterns within a given hexagon, known inflow and outflow between travels zones A and B are proportionally allocated based on the overlap. For example, Hexagon X overlaps with 12\% of Region A, and therefore 12\% of Region A's \emph{outflow} is assigned to Hexagon X. Hexagon Y overlaps with 25\% of Region B, and therefore 25\% of the 12\% outflow is assigned to Hexagon Y as \emph{inflow}.
	}
	\label{figure:commute}
	\end{figure}

 To interpolate the 2017 travel survey network to the hexagonal cells, the homogeneity assumptions of aerial weighting are extended to mobility flows between travel survey zones \cite{Jang2011}. It is assumed that a hexagon cell overlapping with a given origin zone has a proportional quantity of outflow to all its targets. Similarly, hexagon cells overlapping with a given destination zone receive inflow from all relevant origin zones proportional to their intersection with that destination zone. An illustration of the travel flow interpolation to hexagon cells is provided in Figure \ref{figure:commute}. \par

\begin{figure}[h!]

     \includegraphics[width=8cm]{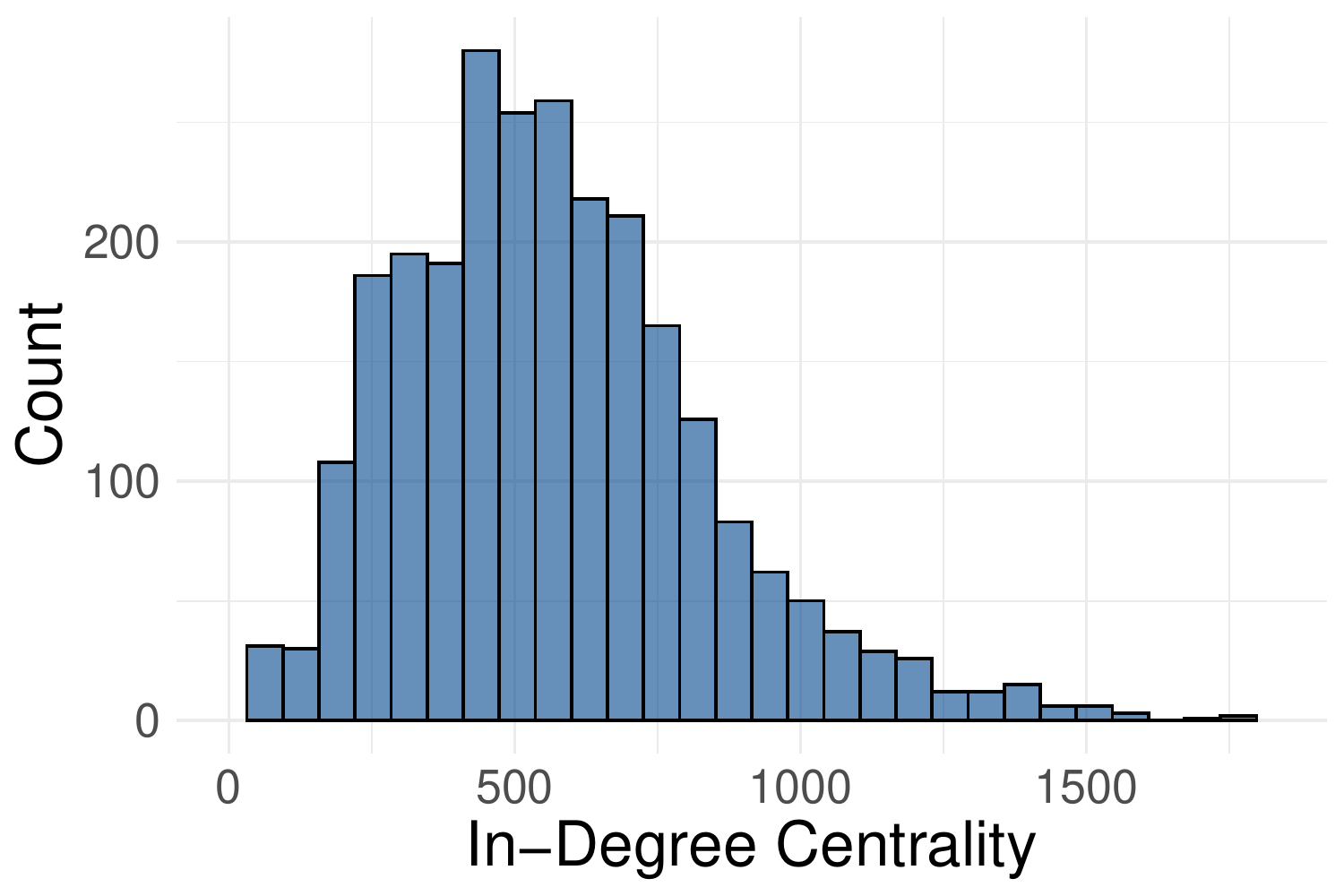}

  \caption{Distribution of in-degree centrality for hexagon cells in the interpolated mobility network.}
  \label{figure:indegree_dist}
\end{figure}

Based on the interpolated mobility network, the in-degree centrality of each hexagon cell is calculated. In-degree centrality is the number of edges that directly flow into a cell, representing the diversity of inflow connections -- associated with a region's time to infection \cite{Hunter2020, central}. The weighted in-degree (total travellers in) and weighted out-degree (total travellers out) are also highly correlated with the in-degree, and have been shown to influence the spread of disease \cite{Francetic10.1093/eurpub/ckab072}. The distribution of in-degree centrality in the hexagon-interpolated mobility network are presented in Figure \ref{figure:indegree_dist} and Table \ref{table:indegree_table}.

\FloatBarrier

\begin{table}[h!]
\begin{tabular}{lr}
    \hline 
{} &    In-Degree Centrality \\
\hline 
mean  &   572.4 \\
std   &   268.6 \\
min   &    83.0 \\
25\%   &   383.0 \\
50\%   &   544.0 \\
75\%   &   719.0 \\
max   &  1829.0 \\
\end{tabular}
\label{table:indegree_table}
\caption{Summary Statistics of in-degree centrality for hexagon cells in the interpolated mobility network.}
\end{table}

\section*{Infection Delay Model} 
The inputs to the Infection Delay Model are the effective distances from all pairs of hexagon cells, calculated under baseline and intervention mobility scenarios -- the baseline being the scenario with no mobility restrictions, i.e.\ no form of lockdown. These effective distances are translated into two sets of infection arrival times, whose differences represent the `infection delay' of an intervention. Subsequently, an algorithm is developed to compute the infection delays of an intervention for a given region over the course of an outbreak, outputting a time series plot defined as the infection delay curve.

\subsection*{Calculating Effective Distance} \label{section:ed}

To calculate the effective distances for the hexagon-scaled mobility network, this analysis uses the `dominant path' effective distance, a metric used in numerous disease arrival time analyses \cite{Iannelli2017, Brockmann2014, Coelho2020, Gautreau2008}, translated into Python by \cite{Iannelli2017}. Measures of the dominant path effective distance focus solely on the most probable path of transmission from hexagon $i$ to $j$. To calculate this value, for every connected origin $i$ and destination $j$ in the network, we calculate the transition rate as the proportion of total travellers beginning in hexagon $i$ who arrive in hexagon $j$, denoted as $ 0 \leq P_{ij} \leq 1$ \cite{Brockmann2014}. The effective distance between hexagons $i$ and $j$ is calculated as:

\begin{equation}
    d_{ij} = 1-\ln({P_{ij}}), 
\end{equation}

which is used as an edge weight for every pair of $i, j$ hexagons, or nodes in the mobility network \cite{Brockmann2014}. These edge weights, greater than or equal to one, are used in a weighted shortest path analysis to determine the dominant path effective distance between every pair of hexagon cells. With $d_{ij}$ calculated for all edges in the network, the dominant path between $i$ and $j$ is chosen as the path which minimizes the sum of effective distance edge weights between them. Finally, the dominant path effective distance between two hexagons ($D_{ij}$) is calculated as the sum of the effective distances along the determined shortest path. This basic dominant path effective distance can be used to detect rankings of arrival times for a given outbreak location \cite{Brockmann2014}. 

The traditional dominant path effective distance model is solely based on the mobility network, captured by $P_{ij}$, and does not have parameters which can incorporate changing epidemiological parameters or rates of mobility reduction in the network. To add epidemiological and mobility-based parameters, useful for a comparative analysis, the effective distance formula is altered to 

\begin{equation}
    d_{ij}= \left(\ln \left( \frac{\beta-\mu}{\kappa}\right) - \lambda\right) -\ln (P_{ij}), 
\end{equation}

shown in \cite{Iannelli2017}, where $\beta$ and $\mu$ are the infection and recovery rate. The mobility compound parameter $\kappa$, representing the proportion of the circulating population, is altered to incorporate mobility reductions, given by $(1.0-\textit{mobility reduction}/100.0)\times \kappa_0$, where the \textit{mobility reduction} goes from $0$ to $100\%$. In this compound parameter, $\kappa_0$ is the mobility rate, chosen to be 10\%, which also ensures the logarithm is positive after the subtraction of $\lambda$, the Euler-Mascheroni constant \cite{Iannelli2017}. As $\kappa_0$ is constant between the baseline and intervention scenarios, its value does not impact the infection delay value when the differences in arrival times are calculated between the two. The reproductive number $R_0$ is chosen to be 2.9, based on an epidemiological characterisation of the MRSP early in the pandemic \cite{DeSouza2020}. The infectious period is chosen to be 9.2 from a mathematical analysis of COVID-19 in Brazil \cite{PintoNeto2021}. The infection rate is thus $R_0/\textit{infectious period}$ = 2.9/9.2, and the recovery rate is given by $1/\textit{infectious period}$=1/9.2.  It is important to note that the transition rate $P_{ij}$ calculation is unaltered from the traditional model. As the mobility compound parameter $\kappa$ rises, $d_{ij}$ decreases, indicating that $i$ and $j$ are effectively farther. Similarly to the traditional model, for every potential outbreak and target hexagon cell in the network,  the dominant path effective distance is generated from the weighted shortest path analysis, generating a $2599 \times  2599$  matrix of effective distances. This method is able to calculate effective distances between hexagons irrespective of whether they are directly or indirectly connected.  \par

Two $2599 \times  2599$ matrices of effective distances are calculated for every potential origin $i$ and destination $j$, under the following mobility flow scenarios:

\begin{enumerate}
    \item No mobility reduction (baseline scenario)
    \item Reduction in mobility based on hexagonal isolation changes 
\end{enumerate}

The first scenario assumes no interventions, where arrival times are calculated using the baseline travel pattern information ($\textit{mobility reduction}=0$). The second scenario assumes that hexagon cells reduce their mobility by the same amount as observed during the first wave of the pandemic, through leveraging the cellphone social isolation information. To determine the extent of the mobility reduction for each region, the marginal change in social isolation from pre-lockdown to post-lockdown is calculated. The initial isolation value for each hexagon is calculated as the mean across March 1 to March 15, the two weeks leading up to the MRSP's lockdown \cite{Siciliano2020}. The lockdown isolation value for each hexagon is calculated as mean from March 16 to March 30 2020, capturing the initial regional responses to lockdown measures. After determining the marginal change in real isolation for each origin hexagon, the effective distance calculation becomes: \par

\begin{equation}
    d^{\textit{intervention}}_{ij}=\ln{\left( \frac{\beta-\mu}{\kappa^{\textit{mobility reduction}}_i} - \lambda \right)} -\ln{(P_{ij})}
    \label{eq:marg}
\end{equation}

This representation of effective distance is used to approximate how rapidly a disease would spread from hexagon $i$ to $j$ given the observed change in pandemic isolation for region $i$. The adjustment of the compound $\kappa$ term to $\kappa^{\textit{mobility reduction}}_i$ is a novel contribution of the study, allowing the analysis to capture heterogeneous changes in mobility based on cellphone mobility data, known to intersect with socio-economic vulnerability in the MRSP \cite{LiSabrinaPereira2020}. 

\subsection*{Infection Delay of Intervention} \label{infection delay}
To generate an estimation of arrival times based on the effective distances, this paper employs the methods used in \cite{Iannelli2017}, dividing the effective distance by the effective velocity, defined as $V^{EF} \approx \beta-\mu$, where $\beta$ is the infection rate and $\mu$ is the recovery rate. The arrival time for a disease to arrive from location $i$ to location $j$, including both the dominant path effective distance $D_{ij}$ (sum of shortest effective distance path from $i$ to $j$) and velocity is thus:

\begin{equation}
    T_{ij} = \frac{D_{ij}}{V^{EF}}
\end{equation}

Having generated the arrival times under both scenarios for every $i, j$ combination, the infection delay by an intervention for an introductory case arriving from origin $i$ to destination $j$ is calculated as:

\begin{equation}
ID_{ij}   = T^{\textit{intervention}}_{ij} - T^{\textit{baseline}}_{ij} 
\end{equation}

The infection delay ($ID_{ij}$) values are calculated for every pair of hexagon cells, generating a $2599 \times 2599$ matrix where each $i, j$ value represents the additional time to a case arriving from $i$ to $j$ given a reduction a traffic proportional to $i$'s real mobility change. \par 

Using known changes in mobility to understand intervention effectiveness takes into account the inequality in regional responses, and allows intervention scenarios to mimic the real capacities of hexagon cells to isolate and adhere to policy guidelines. Having the arrival times in $T^{\textit{intervention}}_{ij}$ reflecting the real mobility changes allows for an infection delay analysis which better captures the lived experience of each of the 2599 hexagon cells in determining the relative benefits from early interventions. \par

From the MRSP's first case of COVID-19 to its widespread presence, this analysis determines the time `added' until a region's first case (infection delay) by an intervention at every hypothetical time $t$, assuming no intervention before $t$. At time $t=0$, only the initial outbreak location $i_0$ has the disease, and each hexagon's infection delay by an intervention is $ID^0_{ij_0}$, representing the change in intervention arrival time relative to the baseline arrival time from $i_0$ to $j$. For every $t \geq 1$, each hexagon cell's infection delay value is determined based on the currently infected regions. To calculate this value, for every hexagon cell $j$ and discrete time step $t$, the following algorithm is developed:
\begin{enumerate}
\item Determine all infected hexagon cells at time $t$ 
\item Determine the infection delay of an intervention across all currently infected hexagon cells relative to destination $j$
\item Select the minimum infection delay value
\end{enumerate}

\begin{figure}[h!]
      \includegraphics[width=12cm]{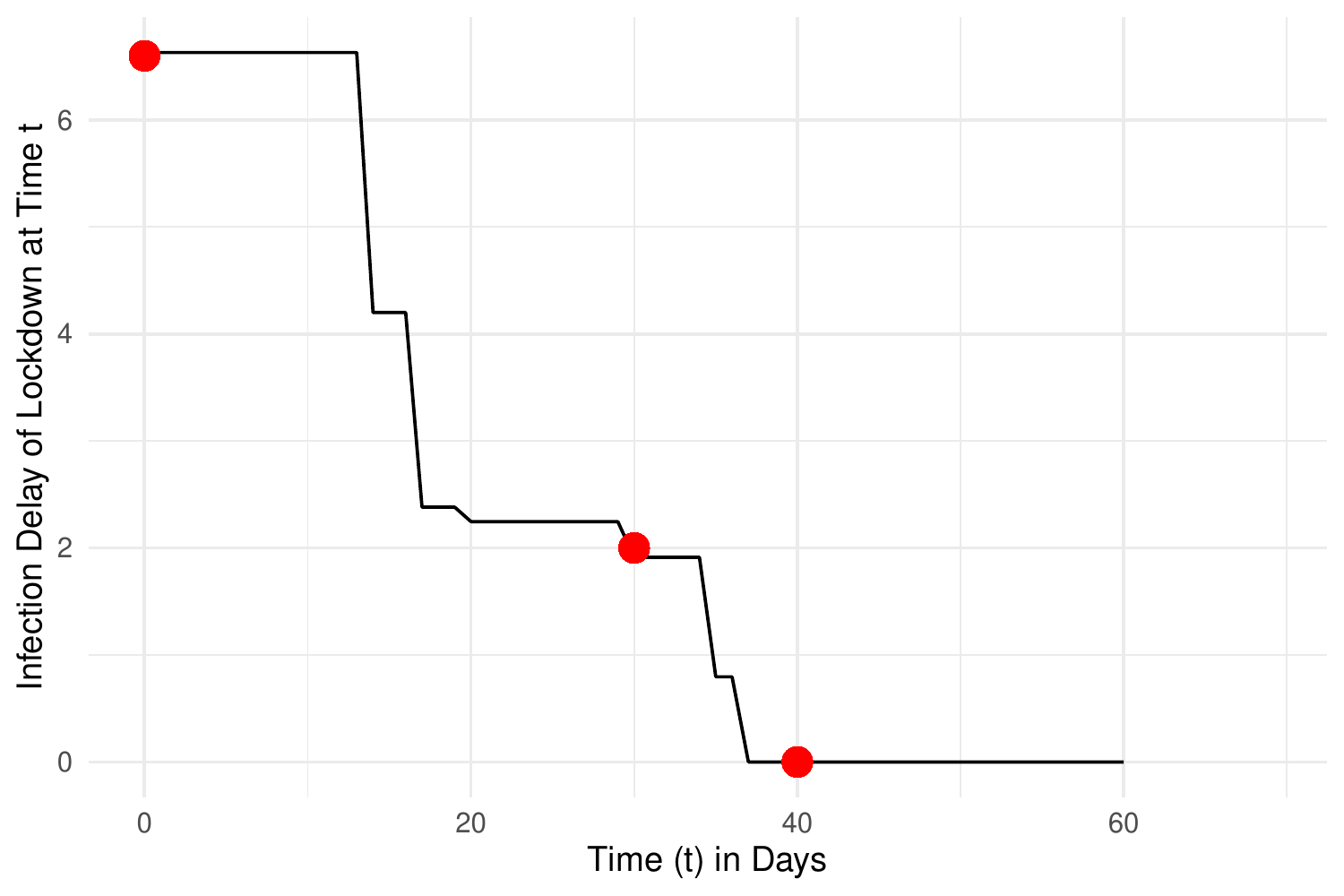}

  \caption{Hypothetical infection delay curve for region-at-risk $A$ caused by a lockdown, following an outbreak beginning in region $B$. At time $t=0$, location $B$ would be the only infected region -- as the outbreak location. At this time, a lockdown would allow region $A$ to gain approximately 6.6 days (y-axis) until its first case of COVID-19. If the disease were to spread unmitigated until time $t=30$ days, a lockdown would provide a gain of only 2 days before region $A$'s first case.  At the 40-day mark following an outbreak in region $B$, without any intervention, region $A$ would already be infected. Thus, a lockdown intervention at this point would have no ability to delay the onset of infection, with a y-axis value of 0. 
}
	\label{figure: example_plot}
	\end{figure}

Following this algorithm, the IDM generates a time-series infection delay curve. An example plot is presented in Figure \ref{figure: example_plot}, for a given hexagon $A$ and outbreak location $B$. There are two primary factors that interact to create the structure of the infection delay curve: (1) the effective distance of infected hexagon cells to the hexagon cell of interest; (2) the degree of mobility reduction of infected hexagon cells. The outbreaks used in this analysis will be simulations calculated from a compartmental epidemiological model.

To generalize the findings of the infection delay analysis to outbreak scenarios other than those observed during COVID-19, epidemic outbreaks are simulated beginning in each of the 2599 hexagons in the MRSP. This paper employs the commuter-oriented susceptible-infected-removed (SIR) model used in \cite{Schlosser2021}, on GitHub as \emph{EpiCommute}. While the original model is used to simulate the spread of COVID-19 in 401 German counties, this analysis uses the 2599 social isolation hexagons, providing their interpolated populations and mobility flows. \par 

For each outbreak scenario, the calculated arrival times are used in conjunction with the IDM to generate infection delay curves for every hexagon cell. The end result is 2598 infection delay curves for every hexagon (excluding its own outbreak), each one encapsulating the infection delay to the first case by an intervention at every time $t$. 

\subsection*{Median Infection Delay Values}
\begin{figure}[!h]%
      \includegraphics[width=12cm]{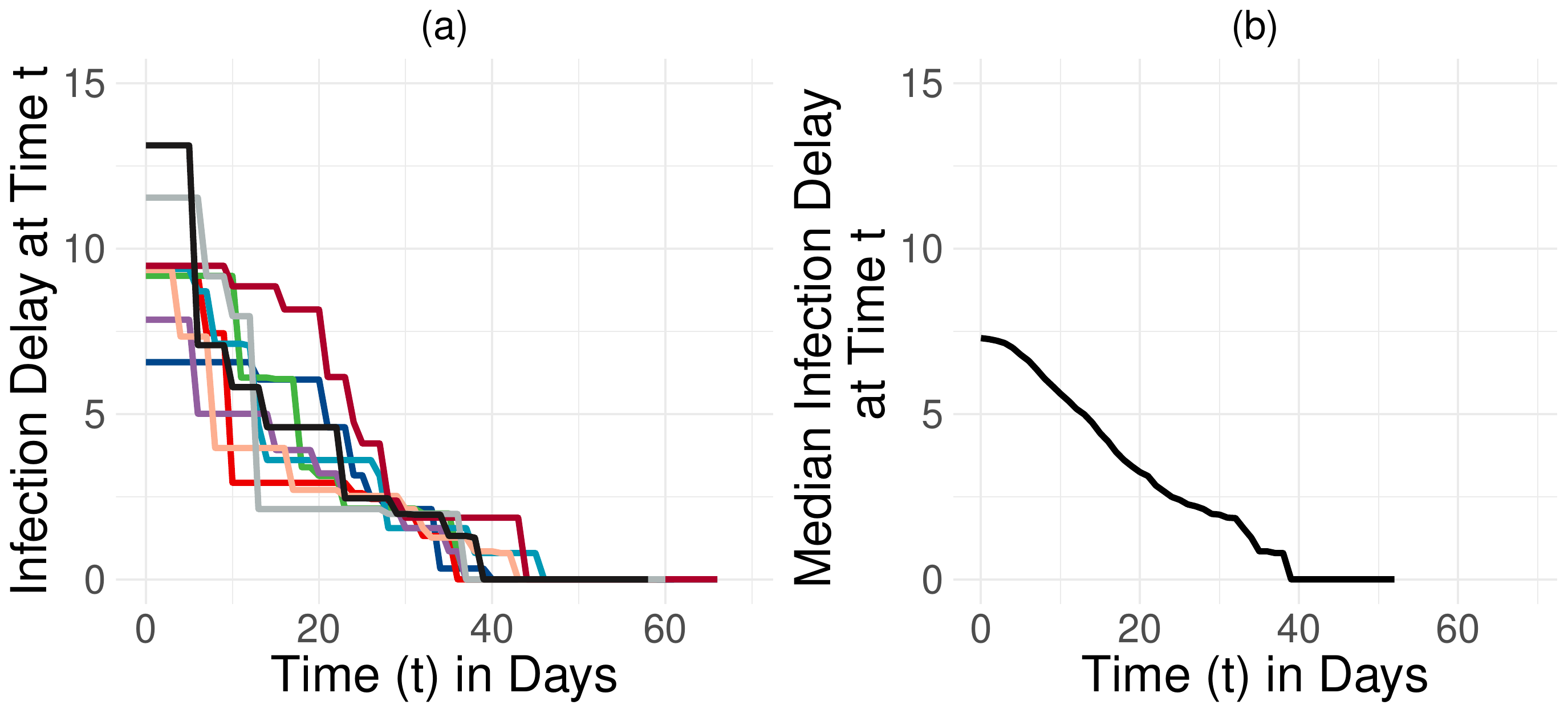}

  \caption{
Illustrative example of infection delay median pipeline for a single hexagon cell, using only 10 outbreaks for visualisation (real analysis uses 2598 outbreak scenarios). From left to right: (a) the infection delay curves are calculated for each outbreak location; (b) the median of those curves are taken at every time $t$ to create a general characterisation of lockdown effectiveness in the region-at-risk.}
	\label{figure:kpipe}
\end{figure}

To extract key information from each hexagon cell's 2598 infection delay curves, the median infection delay from an intervention over the first 10 days is used to summarize the curve. The first 10 days are chosen as they best exemplify the differences in infection delays across early outbreak scenarios, after which the curves begin to converge. Figure \ref{figure:kpipe} displays the pipeline for calculating median infection delay curves for each hexagon cell. In the first set of results, rather than assigning every infection delay curve an equal weight, assuming that each scenario is equally likely, each curve is weighted by the in-degree centrality of its outbreak location. Each hexagon cell is divided into centrality and income quartiles, and their relationships to infection delays are explored. A one-way ANOVA test is performed on the infection delay values to determine whether the differences are statistically significant. In the second analysis, each hexagon cell's 2598 infection delay curves are divided into two groups based on the in-degree centrality of the outbreak location. A student's t-test is performed on the two groups of infection delay values to test whether the differences are statistically significant.

\section*{Results}
\subsection*{Weighted Median Infection Delay Curve}
\begin{figure}[h!]
    \includegraphics[width=12cm]{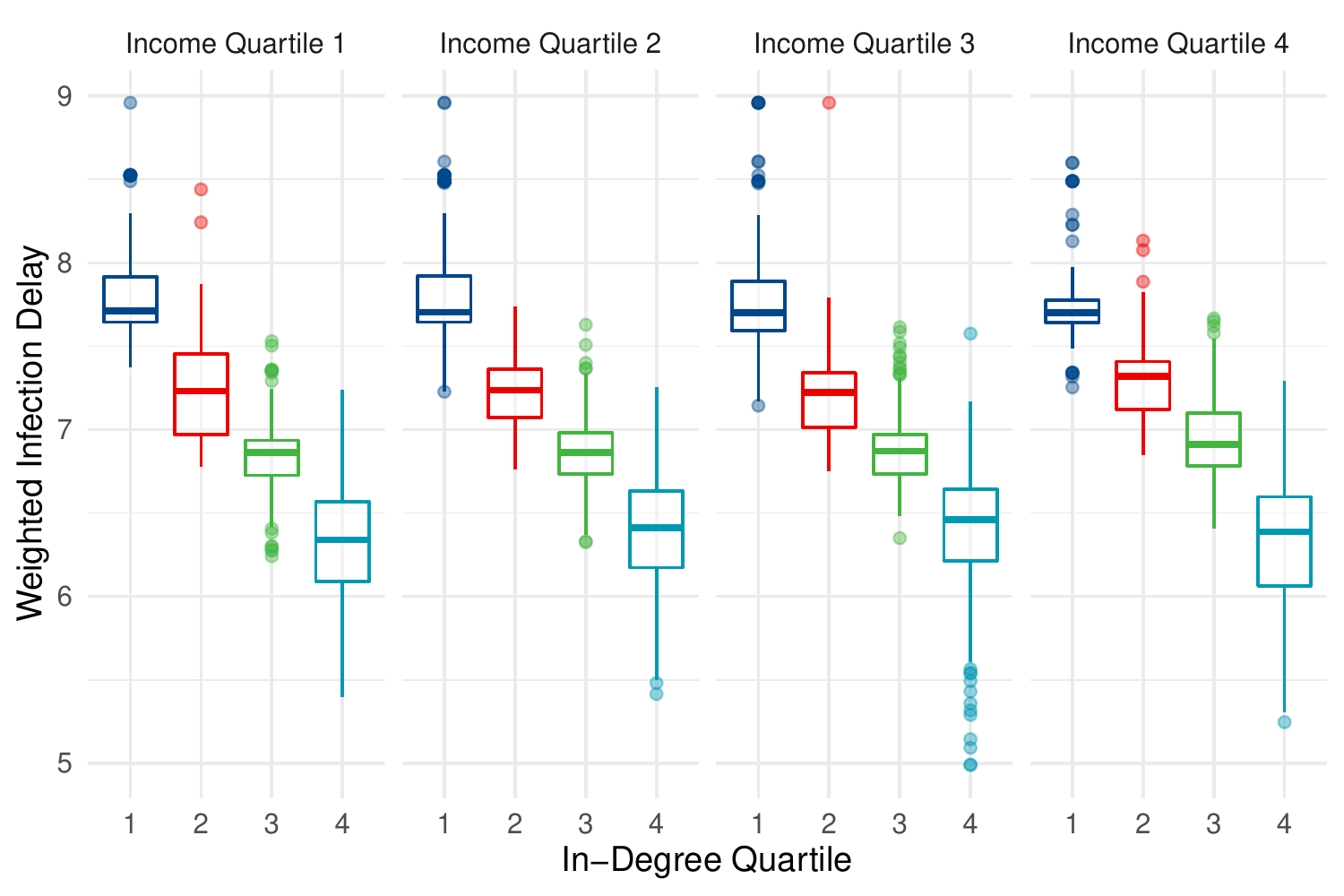}
  \caption{Weighted median infection delay values across in-degree centrality quartiles, while controlling for income.}
\label{fig:income_control}
	\end{figure}

\begin{figure}[h!]
      \includegraphics[width=10cm]{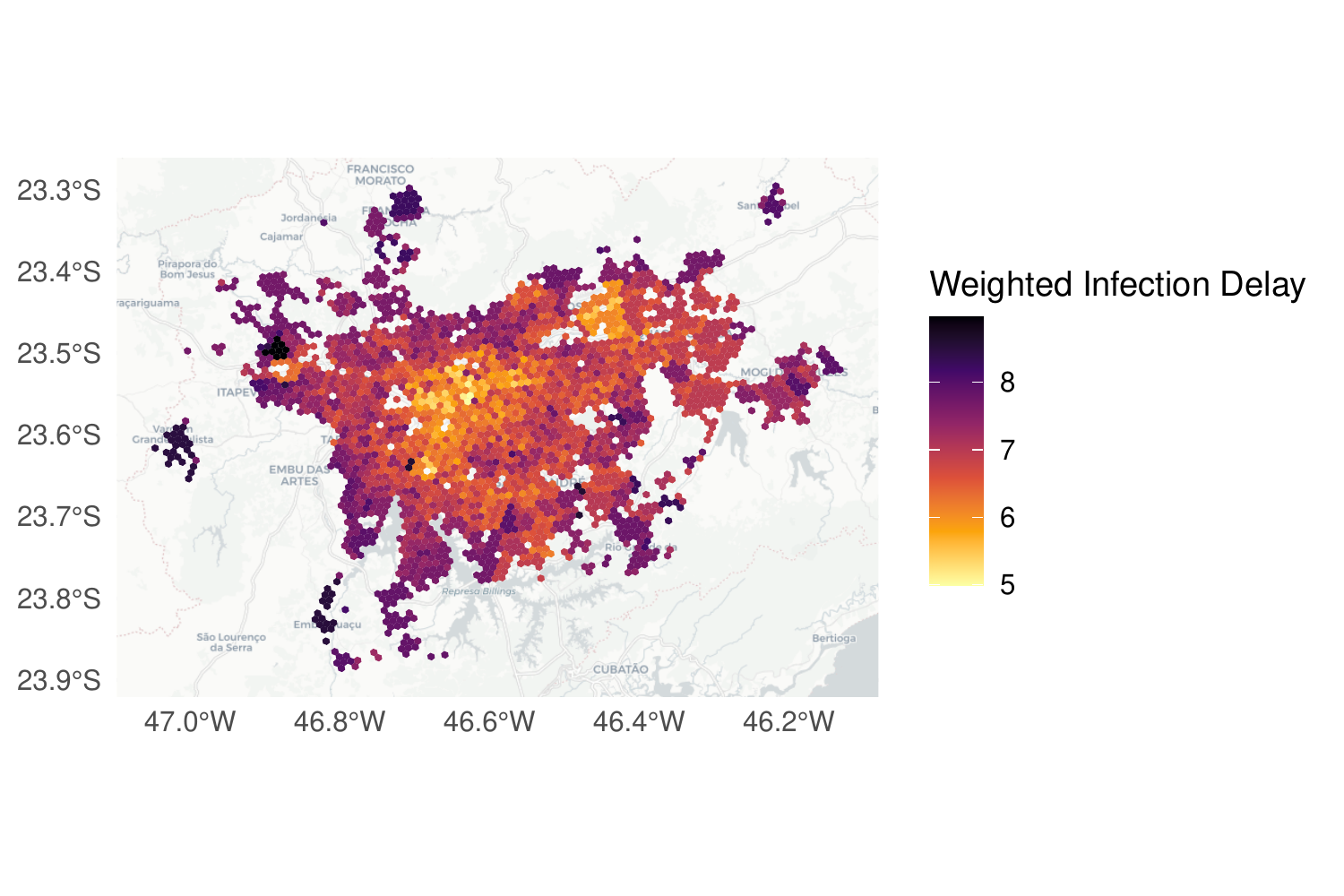}
  \caption{Geographic distribution of weighted median infection delay values over the first ten days of an outbreak.}
	\label{fig:ID_map}
	\end{figure}

The relationship between greater centrality and lower infection delay values is displayed in Figure \ref{fig:income_control} and Table \ref{tab:income_control}. Within every income quartile, greater centrality is associated with a lower median infection delay value. These differences between infection delay values across centrality quartiles, controlling for income quartile, are statistically significant at the $p<0.01$ level based on the one-way ANOVA test. Figure \ref{fig:ID_map} shows the geographic distribution of weighted median infection delay values.     \par

\FloatBarrier

\begin{table}[h!]
    \begin{tabular}{llr}
    \hline
Income Quartile & In-Degree Quartile &  Weighted Median Infection Delay \\
(Median BRL) & (Median In-Degree)   \\
        \hline

         1  (937.33) &            1 (268) &   7.71 \\
          &            2 (452) &   7.23\\
          &            3 (625) &   6.86\\
          &            4 (813) &   6.34\\
          \hline
         2 (1212.41) &            1 (268) &   7.70 \\
          &            2 (468) &   7.23\\
          &            3 (623) &   6.86\\
          &            4 (791) &   6.41\\
           \hline
         3 (1712.48) &            1 (268) &   7.70 \\
          &            2 (470) &   7.22\\
          &            3 (629) &   6.87\\
          &            4 (847) &   6.46\\
           \hline
         4 (3483.72) &            1 (325) &   7.70 \\
          &            2 (469) &   7.31\\
          &            3 (635) &   6.91\\
          &            4 (944) &   6.38\\
\end{tabular}
    \caption{Weighted median infection delay values across income and in-degree quartiles. Median income per capita (Brazilian Real per Month) and in-degree centrality within each quartile subgroup is shown. The differences in weighted median infection delay values across in-degree centrality quartiles are statistically significant ($p<0.01$).}
    
    \label{tab:income_control}
\end{table}

    \FloatBarrier

\begin{figure}[h!]
       \includegraphics[width=12cm]{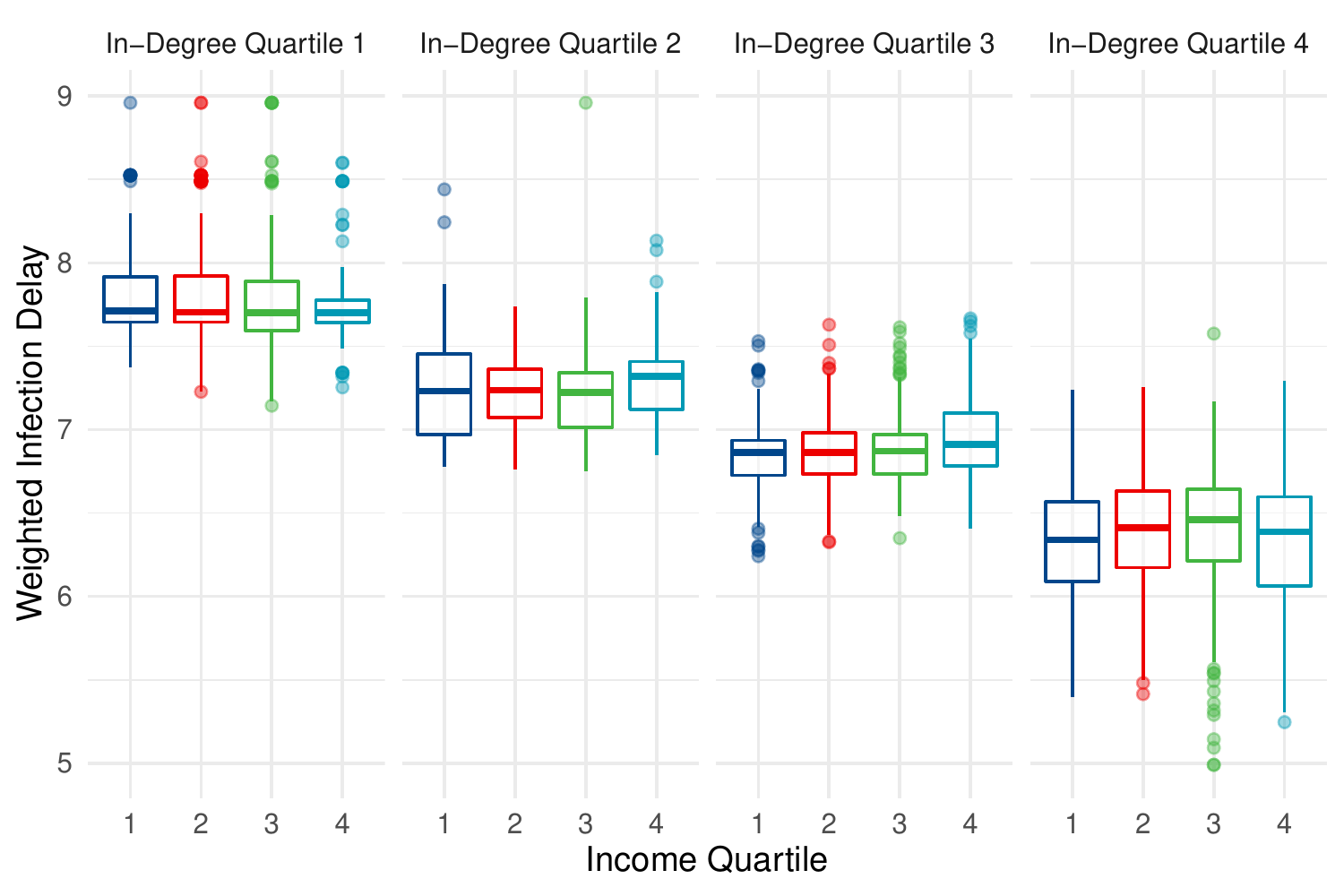}

  \caption{Weighted median infection delay values across income quartiles, while controlling for in-degree centrality.}
	\label{fig:centrality_control}
	\end{figure}

Figure \ref{fig:centrality_control} displays the distribution of infection delay curves across income groups, controlling for their levels of centrality. Observing the hexagon cells' infection delays from Figure \ref{fig:centrality_control}, this analysis finds no discernable trend across income groups. The median infection delay values of hexagon cells in the bottom 25\% of centrality are between 7.5 and 8 days. Hexagon cells in the highest centrality quartile all have median infection delay values between 6 and 6.5 days.

	   \FloatBarrier

\subsection*{Division by Outbreak Location Centrality}
\begin{figure}[h!]
        \includegraphics[width=12cm]{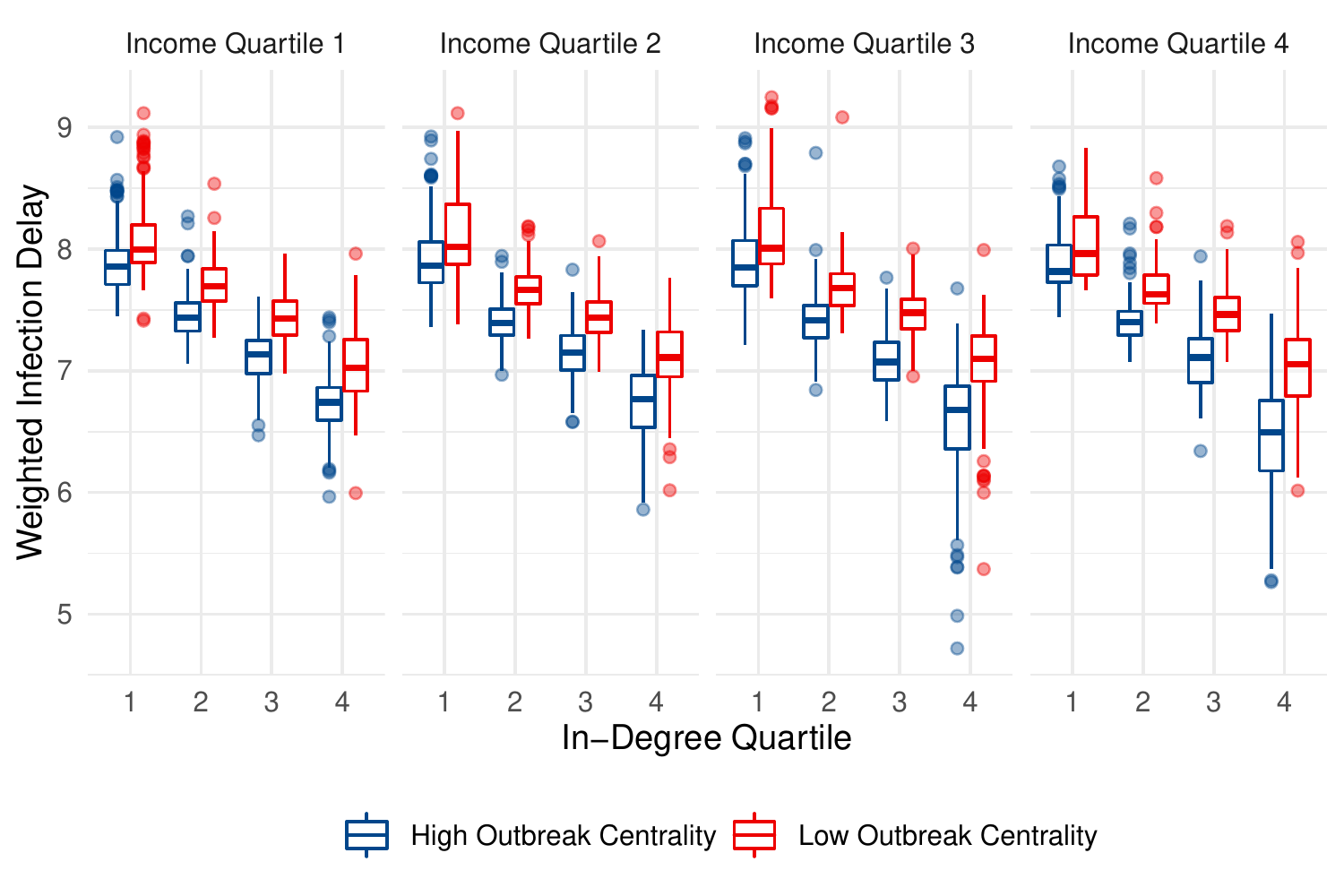}

  \caption{Un-weighted infection delay values across in-degree centrality quartiles, while controlling for income. Each region's infection delay values are calculated and displayed for outbreak scenarios in the upper and lower 50\% of in-degree centrality.}
	\label{fig:outbreak_income_control}
\end{figure}

\begin{figure}[h!]
         \includegraphics[width=12cm]{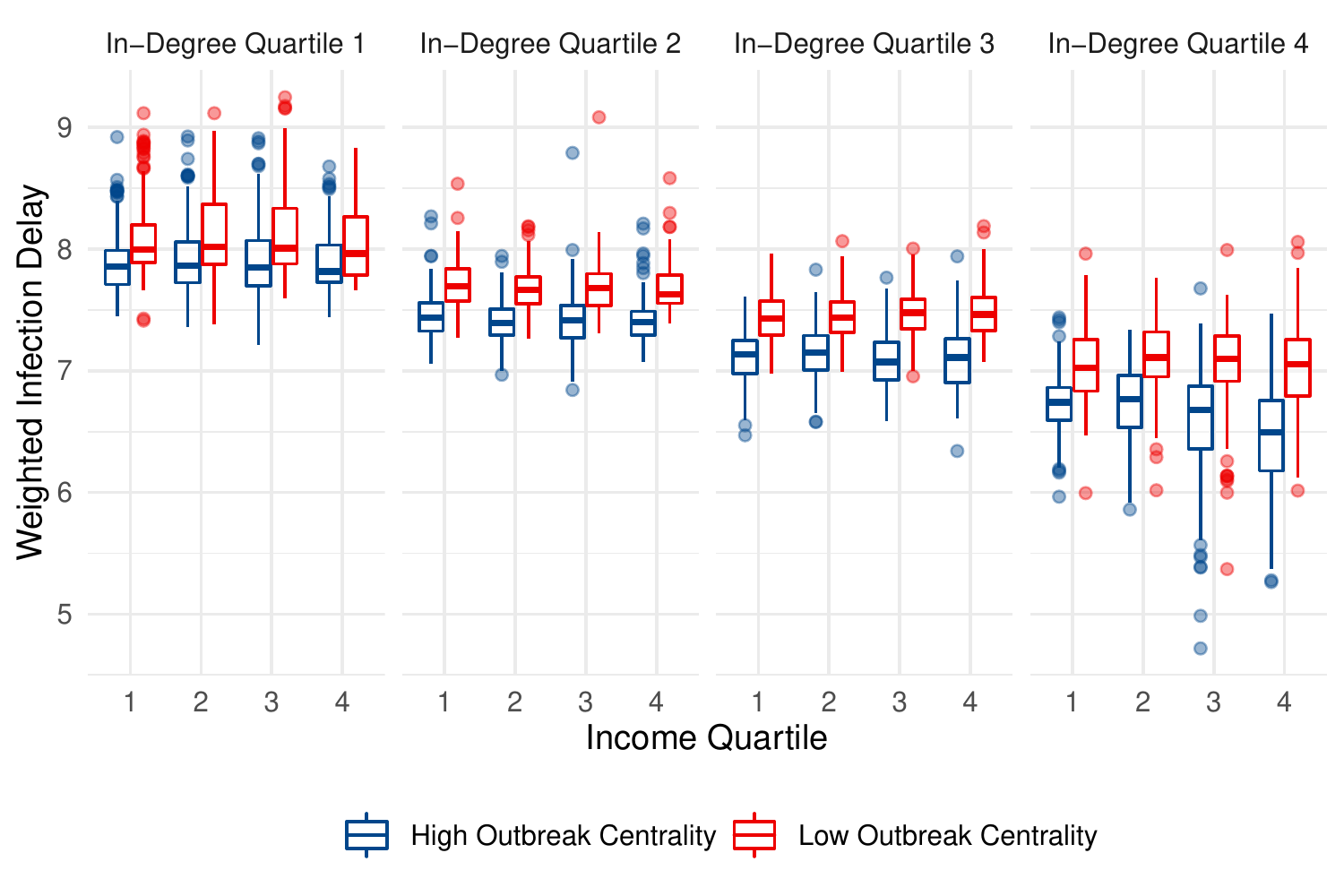}

\caption{Un-weighted infection delay values across income centrality quartiles, while controlling for in-degree centrality. Each region's infection delay values are calculated and displayed for outbreak scenarios in the upper and lower 50\% of in-degree centrality.}
    \label{fig:outbreak_centrality_control}
	\end{figure}

Each hexagon cell's infection delay value is subsequently calculated and shown when the outbreak location is in the bottom versus top 50\% of centrality.  For every hexagon cell, this creates two infection delay values, shown side-by-side in Figures \ref{fig:outbreak_income_control} and \ref{fig:outbreak_centrality_control}. We see that greater centrality is associated with lower infection delays, irrespective of income, and no clear pattern across income groups is observed when controlling for centrality -- similarly to Figures \ref{fig:income_control} and \ref{fig:centrality_control}. The results also show that irrespective of the income and centrality grouping, outbreaks beginning in hexagon cells of lower centrality lead to greater infection delays of lockdowns. The student's t-test indicates a statistically significant ($p<0.01$) difference between infection delay values depending on whether the outbreak location's centrality is below of above the median.

	   \FloatBarrier

		   \FloatBarrier

\section*{Discussion}
This analysis has sought to uncover how the socio-economic and network characteristics of a region relate to the delay of its first case from an early intervention. The results of the Infection Delay Model indicate that the centrality of a region, independent of its income level, plays the largest role in determining how an early intervention will delay their first infection. There is no discernable relationship between income levels and the ability of a lockdown to slow the arrival of disease when controlling for centrality. Although previous studies have shown that vulnerable communities with lower isolation levels have higher infection rates of COVID-19 \cite{Lee2021, LiSabrinaPereira2020, Cordes2020}, our results suggest that the influence of socioeconomic and isolation inequalities in determining disease arrival is overridden by the outsized influence of centrality in the network. As an effective distance-based analysis, more central regions tend, on average, to be `closer' to infected regions. This proximity reduces the potential infection delay of a lockdown, with an opposite mechanism in play for less central regions.

The literature produced during the COVID-19 pandemic has thoroughly highlighted the importance of socio-economic factors and their relationship to disease risk, rationalizing their use as more than a passive add-on to network-based results. The growing prioritization of socio-economic inequalities as a driving force of disease risk is exemplified in studies such as \cite{Sheen2020}, who study how eviction rates in Philadelphia have a measurable impact on the spread of COVID-19. The Infection Delay Model reflects socio-economic inequalities in the MRSP by incorporating real-life mobility reductions -- known to be weaker in vulnerable areas \cite{LiSabrinaPereira2020} -- as a core component in the effective distance network analysis. Income is then used as a key axis to explore infection delays, found to be overpowered by a region's centrality.

Rather than contradicting existing literature on the health burden inequalities associated with socio-economic status, this paper uncovers an unexplored perspective on pandemic preparedness. The emphasis of previous literature on case, death, and hospitalization counts illuminate how vulnerable groups are most at risk during the course of an outbreak \cite{LiSabrinaPereira2020, Rocha2021, Jay2020, Lee2021, Cordes2020, Pereira2021, Coelho2020}. This paper targets a different, intervention-focused question: \emph{How much time can be gained to a region's first case from an early lockdown?} It cannot be assumed that the same mechanisms leading to greater disease risk \emph{during} an outbreak lead to reduced intervention effectiveness \emph{prior} to an outbreak. Our results, in conjunction with the established literature on socio-economic vulnerability and COVID-19, illuminate an additional burden faced by low-income, centrally located regions.

A  major contribution of this study is its generalized, forward-looking characterisation of intervention effectiveness. Rather than relying on a single set of initial conditions when modelling a disease, or using a subset of transport hubs as outbreak locations, this analysis incorporates \emph{all} possible outbreak locations when assessing how early interventions lead to infection delays. This allows for broad understandings of intervention effectiveness whose validity is not reliant on the next epidemic beginning in the same location as the last. This addresses the recently explored importance of outbreak locations on disease trajectories, providing generalizable insights for future disease preparedness \cite{Schlosser2021}. We are able to use the abundance of scenarios to generate weighted median infection delay values (Figures \ref{fig:income_control} and \ref{fig:centrality_control}), emphasizing the dominant role of centrality. Further, we can divide outbreak locations into low and high centrality groups (Figures \ref{fig:outbreak_income_control} and \ref{fig:outbreak_centrality_control}), and show that the infection delays of interventions vary based on the centrality of the outbreak location. We see that irrespective of the income or centrality quartile of recipient regions, outbreaks beginning in less central regions tend to lead to greater slowdowns.

Research into the effectiveness of government interventions to slow disease spread is essential, as the disaster resulting from the COVID-19 pandemic and its emerging new variants continues globally. The novel Infection Delay Model proposed in this study provides a method of capturing how mobility reductions can slow the spread of an outbreak while considering the network patterns of mobility flows, an important element of intervention effectiveness. The data-linkage approach, interpolating travel behaviour and socio-economic data, allowed for insights into the social context of regions and how interventions can delay a region's first case. The unique integration of cellphone mobility data into the effective distance metrics has captured heterogeneous changes in isolation, found in prior literature to intersect with socio-economic inequalities \cite{LiSabrinaPereira2020, Lee2021}. While this analysis is focused on Brazil, a region where income, health, and transport inequalities are stark \cite{Malta2020}, the presented approach can be applied in other regions to observe the intersection of intervention effectiveness, centrality, and socio-economic vulnerability. Similarly, the epidemiological parameters in this analysis are chosen to mimic COVID-19, but a novel variant or disease's reproduction rate and infectious period could be used as substitutes. Adopting interdisciplinary methodologies to investigate the effectiveness of interventions, with a focus on exploring inequalities, may provide novel insights into the factors driving the unequal playing field exposed during the COVID-19 pandemic.

\subsection*{Limitations and Future Directions}

Based on the Infection Delay Model algorithm, the delays calculated for a given region are dependent on the reductions of mobility flows that arrive to it, rather than its own mobility reduction. This operates well under a regime where first cases arrive from individuals travelling from other locations. Advancements of the Infection Delay Model which capture how a first disease introduction to a region can originate from one of its residents travelling elsewhere would capture an important dimension of disease transmission. This may lead socioeconomic and isolation inequalities to play a stronger role in shaping infection delay curves. Further, rather than calculating the time to a region's first case, a case threshold such as 5\% infection-rate of the population could be implemented, in which case a region's own social isolation capabilities would more directly impact its infection delay value. These adaptations of the Infection Delay Model can expand its scope in capturing the concept of intervention effectiveness, as its current focus on the delay to a region's first case is only one important element.

When considering cellphone data sources, originally collected for commercial purposes, coverage bias should be noted. As a cellphone analytics company, the sample of users in the Inloco/Incognia data set is determined by their market share, rather than an emphasis on representative samples \cite{Tizzoni2014}. The near-global ubiquity of cellphones does not preclude biases, as possession and use rates vary across demographic and income groups \cite{Kraemer2020}. The elderly are often underrepresented in such samples, while educated urban males are overrepresented relative to lower-income individuals \cite{Kraemer2020}.

In a preliminary analysis, a modified radiation model was used to determine if the results using real commuting data could be replicated with a generalized model. When observing the outbreak locations which led to above and below average infection delays for the rest of the MRSP, the radiation network overstated the influence of income-related mobility reductions relative to centrality. This may have occurred because the radiation model failed to replicate regional hubs with disproportionately large connectivity throughout the commuting network. This caveat should be considered for future research using effective distance-based metrics on artificially generated commuting data.

The suitability of integrating traditional household travel survey data with the aggregated social isolation cellphone data deserves exploration by future research. This study recommends comparing granular cellphone mobility location pairs, and observing how daily travel patterns and their changes after lockdown resemble those found in this paper's analysis. If providing similar results, the greater anonymity of the aggregated social isolation data may render it a more readily accessible and minimally invasive method for granular mobility-related studies.


\begin{backmatter}

\section*{Abbreviations}
\begin{itemize}
    \item NPI: Non-pharmaceutical intervention
    \item MRSP: Metropolitan Region of São Paulo
    \item IDM: Infection Delay Model
\end{itemize}

\section*{Availability of Data and Materials}
The datasets and code supporting the conclusions of this article are both openly available. Data sets are stored on Zenodo repository here: \url{https://doi.org/10.5281/zenodo.5947174}. The code is stored as a GitHub repository (Python/R) here: \url{https://github.com/shivyucel/infection-delay-project}, and archived at time of submission as a Zenodo repository here \url{https://doi.org/10.5281/zenodo.6008499}.

\section*{Competing interests}
The authors declare that they have no competing interests.

\section*{Funding statement}
P.S.P. was supported by the São Paulo Research Foundation (FAPESP) under grant number 16/18445-7

\section*{Author's contributions}
All authors conceived and designed the study. S.Y. implemented the models, carried out the analysis and wrote the first draft. R.H.M.P. and P.S.P. performed data curation and contributed with the formal analysis. C.Q.C. performed the initial analyses and coordinated the project. All authors discussed, edited, and reviewed the manuscript, and gave final approval for publication.

\section*{Acknowledgements}
We would like to thank Inloco/Incognia, for providing access to valuable privately owned mobility data.

\bibliographystyle{bmc-mathphys} 
\bibliography{bmc_article}      





\end{backmatter}
\end{document}